\documentclass[twocolumn, letter]{jpsj2}

\def\runauthor{Y. YOSHIDA {\it et al.}}
\usepackage{bm}

\title{Specific Heat Study on a Novel Spin-Gapped System : (CH$_3$)$_2$NH$_2$CuCl$_3$ }

\author{Yasuo YOSHIDA\thanks{E-mail address: yasuotap@mbox.nc.kyushu-u.ac.jp}, Osamu WADA, Yuji INAGAKI,$^{1}$ Takayuki ASANO,$^{2,4}$\\ 
Kenji TAKEO,$^{2}$ Tatsuya KAWAE, Kazuyoshi TAKEDA, and Yoshitami AJIRO$^{3,4}$}

\inst{Department of Applied Quantum Physics, Kyushu University, Fukuoka 812-8581, Japan\\
$^{1}$ Department of Chemistry, Kyushu University, Fukuoka 812-8581, Japan\\
$^{2}$ Department of Physics, Kyushu University, Fukuoka 812-8581, Japan\\
$^{3}$ Department of Chemistry, Kyoto University, Kyoto 606-8502, Japan\\
$^{4}$ CREST Japan Science and Technology Corporation (JST), Saitama 332-0012, Japan}

\recdate{\today}

\abst{Specific heat measurements down to 120mK have been performed on a quasi-one-dimensional $S=1/2$ 
spin-gapped system (CH$_3$)$_2$NH$_2$CuCl$_3$ in a magnetic field up to 8 T. 
This compound has a characteristic magnetization curve which shows a gapless ground state and a plateau at 1/2 of the saturation value.   
We have observed a spontaneous antiferromagnetic ordering and a field-induced one below and above the 1/2 plateau field range, respectively. 
The field versus temperature phase diagram is quite unusual and completely different from those of the other quantum spin systems investigated so far.
In the plateau field range, a double-structure in the specific heat is observed, reflecting the coexistence of ferromagnetic and antiferromagnetic excitations.
These behaviors are discussed on the basis of a recently proposed novel quantum spin chain model consisting of 
weakly coupled ferromagnetic and antiferromagnetic dimers.}

\kword{specific heat, spin-gapped system, field-induced magnetic ordering}

\begin{document}

\maketitle
In recent years, field-induced quantum phenomena in spin-gapped systems, such as 
field-induced magnetic orderings (FIMO) and plateaux in the magnetization curve $M$($H$), 
have been brought to attention as one of the hottest topics in condensed matter physics. 
Spin-gapped systems are magnetically inactive and have energy gaps 
in the energy spectrum at zero or finite fields because of forming antiferromagnetic singlet dimers, 
but magnetic excited states exist above these energy gaps. 
Thus, by application of an external field, 
field-induced magnetic states are realized due to the Zeeman energy gain above certain critical fields. 
In these magnetic states, spin components perpendicular to the field direction show three-dimensional (3D) long-range 
orderings triggered by interchain and/or interdimer interactions at low enough temperatures, 
and the resultant ordered phase ususally 
has the reentrant $H$-$T$ phase boundary with a parabolic shape between two critical fields. 
This kind of the FIMO was firstly investigated in an alternating chain system Cu(NO$_3$)$_2$$\cdot$2.5H$_2$O 
\cite{cunitrate} and explained by Tachiki and Yamada based on the mean-field theory. \cite{TM} 
Recently, Nikuni {\it et al.} have discussed this kind of the FIMO 
on the basis of a Bose-Einstein condensation of excited magnons, \cite{nikuni} and then
FIMO observed in spin-gapped compounds have been analyzed within this framework. \cite{Tl, K, NDMAP} 

Our model substance in this study is an $S=1/2$ quasi one-dimensional (1D) spin-gapped system 
(CH$_3$)$_2$NH$_2$CuCl$_3$ (DMACuCl$_3$). 
This compound has a unique magnetic chain forming an alternating chain of antiferromagnetic ($S=0$) 
and ferromagnetic ($S=1$) dimers coupled by an intervening weak interaction, 
as proposed recently from the low-$T$ x-ray diffraction and magnetization measurements. \cite{watson, ina} 
In this system, 
there are three kinds of nearest-neighbor intrachain  interactions, 
$J_\mathrm{AF}$, $J_\mathrm{F}$ and $J$, 
in a sequence of - $J$ - $J_\mathrm{AF}$ - $J$ - $J_\mathrm{F}$ - $J$ -, 
where $J_\mathrm{AF}$ and $J_\mathrm{F}$ 
are antiferromagnetic (AF) and ferromagnetic (F) interactions responsible 
to the AF and F dimers, respectively, 
and $J$ is an intervening weak interaction, 
$\mid J\mid < \mid J_\mathrm{AF}\mid $ and $\mid J_\mathrm{F}\mid $. 
From this model, 
it is naturally expected that the system behaves as the isolated mixed F and AF dimer model 
at $k_\mathrm{B}T>\mid J\mid$ 
as the specific heat in zero field and the magnetic susceptibility 
are well described by the simple sum of F and AF dimers. \cite{ajiro} 
In fields, however, 
a crucial role of the intervening interaction $J$ begins to surfice as 
was demonstrated in the magnetization measurement. \cite{ina} 
In Fig. 1, 
$M$($H$) initially increases rapidly and reaches to 
a half value of the saturated magnetization 
$M_\mathrm{S}$, at $H_\mathrm{C1}=2.0$ T. 
Then it takes a constant value of $M=1/2M_\mathrm{S}$ 
(a 1/2 magnetization plateau) 
up to about $H_\mathrm{C2}=3.5$ T followed by a gradual increase 
to the saturation at $H_\mathrm{S}=14$ T. \cite{ina} 
The important point to be noted is that the observed peculiar 
$M$($H$) can be explainable only 
by considering sizable interactions between F and AF dimers. 
Because otherwise the magnetization of isolated F dimers 
would abruptly jump to a partially polarized 1/2 plateau state 
at the infinitesimally weak field and also 
the magnetization of isolated AF dimers would abruptly jump to saturate 
at $H_\mathrm{C}=J_\mathrm{AF}/g\mu_\mathrm{B}$. 
The initial increase up to $M=1/2M_\mathrm{S}$ is attributable to the magnetization process of the F dimers 
interacted through the effective weak AF interaction mediated by intermediate singlet AF-dimers. 
On the other hand, the gradual increase above $H_\mathrm{C2}$ is attributable to the magnetization process of the 
AF dimers interacted through the interaction mediated by intermediate triplet F dimers. 
More importantly, 
from the above scenario the system is 
in a novel quantum spin state with coexistence of the gapless 
F and gapped AF excitations. 
As a consequence, 
the system does not have an energy gap at zero field but 
a gap is opened to exhibit the magnetization plateau at $H_\mathrm{C1}$. 
With further increasing field, the gap is closed at $H_\mathrm{C2}$ and 
there appears the field-induced magnetic phase 
which is stable over a wide field range between $H_\mathrm{C2}$ 
and $H_\mathrm{S}$. 

\begin{figure}[b]
\begin{center}
\includegraphics*[width=7cm]{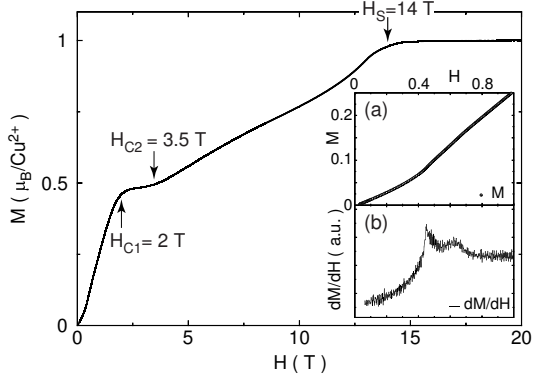}
\caption{The magnetization curve $M$($H$) of DMACuCl$_3$ in the field-range up to 20 T at 500 mK. \cite{ina} 
A 1/2 plateau is observed in the field range of 2 T $\leq H\leq$ 3.5 T.
Insets: The enlarged plots in low fields. (a) $M$($H$), and (b) d$M$/d$H$ versus $H$. 
The anomaly at 0.45 T shows the possible spin-flop transition, 
indicating that the system is in the antiferromagnetic ordered state below the 1/2 plateau field range. }
\end{center}
\end{figure}

In this letter, 
we report the result of the specific heat measurement on DMACuCl$_3$ in a magnetic field up to 8 T. 
The specific heat measurement provides 
the most reliable way to detect a phase transition, 
and the nature of low-lying excitations 
is revealed possibly from temperature dependence of the specific heat. 
In the 1/2 plateau field range, 
we observed the characteristic double-structure in temperature dependence of the magnetic specific heat $C_\mathrm{m}(T)$ 
which reflects the dual character of the low-lying excitations due to coexistence of F and AF dimers. 
Further, unconventional $H$-$T$ phase boundaries are revealed 
for the spontaneous AF ordering and FIMO below and above the plateau field range, respectively. 
These are completely different from those in 
many spin-gapped compounds reported so far. \cite{Tl,K,NDMAP,ore}
Powder samples of DMACuCl$_3$ were prepared by the slow evaporation method. \cite{roger} 
Specific heat measurements were performed by the adiabatic heat-pulse method using a $^3$He-$^4$He dilution refrigerator 
and 8 T superconducting magnet. 



Magnetic field effects on $C_\mathrm{m}(T)$ in three ranges, 
$H<H_\mathrm{C1}$, $H_\mathrm{C1}<H<H_\mathrm{C2}$ and $H>H_\mathrm{C2}$, 
are shown in Figs. 2(a), (b) and (c), respectively. 
Fig 2(b) includes the data at 4 T ($>H_\mathrm{C2}$) for an easy comparison.  
To estimate $C_\mathrm{m}(T)$ from the total specific heat $C_\mathrm{p}(T)$, 
we subtracted the lattice contribution $C_\mathrm{l}(T)$ which was 
determined in the previous work at higher temperatures as shown in the inset of Fig. 2(a). \cite{ajiro} 
 
\begin{figure}[bp]
\begin{center}
\includegraphics*[width=8cm]{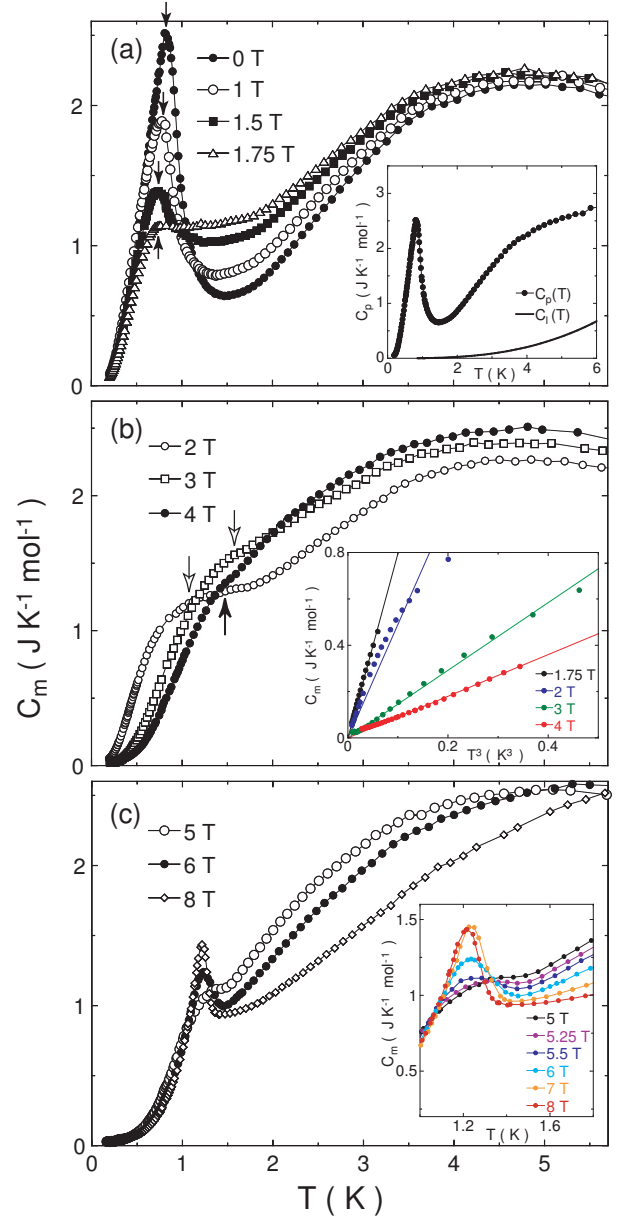}
\caption{$C_\mathrm{m}(T)$ in magnetic fields. 
(a) The data for $H<H_\mathrm{C1}$. 
Arrows indicate the peak temperatures of the 1st peak in magnetic fields.  
The inset of (a): The filled circle and the solid curve indicates $C_\mathrm{p}(T)$ at zero field 
and the lattice contribution $C_\mathrm{l}(T)$, respectively.
(b) The data for $H_\mathrm{C1}<H<H_\mathrm{C2}$ and 4 T ($>H_\mathrm{C2}$).  
Down arrows indicate temperatures of edges of the shoulder at 2 and 3 T, and an up arrow that at 4 T. 
The inset of (b): Field dependence of the $C_\mathrm{m}$ as a function of $T^3$.  
(c) The data for $H>H_\mathrm{C2}$. 
The inset of (c): The low-$T$ $C_\mathrm{m}(T)$ in the field-range between 5 T and 8 T.}
\end{center}
\end{figure}

$C_\mathrm{m}(T)$ below $H_\mathrm{C1}$ is shown in Fig. 2(a). 
At zero field, the sharp peak due to the spontaneous AF ordering (hereafter referred to as a 1st peak) 
and a broad maximum were observed at 800 mK and 4.5 K, respectively in accordance with the previous work. \cite{ajiro} 
With increasing field, the 1st peak collapses gradually into a small hump and finally into a shoulder 
shifting to lower temperatures, which suggests suppression of the spontaneous ordering by application of the external field. 
Simultaneously, a local minimum at around 1.5 K diminishes in depth. 
Moreover, the broad maximum at 4.5 K is enhanced and shifts slightly to low temperatures depending on the field-strength. 

In the 1/2 plateau field range $H_\mathrm{C1}<H<H_\mathrm{C2}$, the 1st peak disappears completely as shown in Fig. 2(b). 
Instead, a shoulder in $C_\mathrm{m}(T)$ with a low-$T$ exponential behavior is observed 
in addition to the broad maximum. 
At 2 T, the shoulder and the broad maximum are observed at around 1 K and 4.5 K, respectively. 
With increasing field to 3 T, both characteristic temperatures shift to the higher and lower temperatures, 
merging further with each other. 
This remarkable feature of the double-structure in $C_\mathrm{m}(T)$ (the shoulder and broad maximum) is also 
characteristic of the quantum ferrimagnetic chains. \cite{ferri2} 
This point will be discussed in the final part of this letter. 

At 4 T ($>H_\mathrm{C2}$), 
the shoulder begins to shift to lower temperatures 
with decreasing its height. 
This field dependence of the shoulder from 3 T to 4 T 
is opposite to that from 2 T to 3 T.  
Accordingly, it is natural that the shoulder at 4 T is caused by another origin.   
This can be confirmed by a low-$T$ behavior in $C_\mathrm{m}(T)$. 
In general, when an anomaly in $C_\mathrm{m}(T)$ is due to the 3D long-range AF ordering, 
the $T^3$ dependence should hold in the temperature range up to about 2/3 of the critical temperature 
due to the AF spin-wave excitation from the N\'eel state.   
As shown in the inset of Fig. 2(b), 
the $T^3$ dependence clearly holds at 1.75 T ($<H_\mathrm{C1}$) and 
4 T ($>H_\mathrm{C2}$), 
while it is not seen at 2 T and 3 T in the 1/2 plateau field range. 
Namely, 
the shoulder at 4 T is an anomaly due to the FIMO 
in contrast with that at 2 T and 3 T. 
More interestingly, 
these behaviors apparently indicates that  
the ground state of the system changes in the course of 
gapless - gapped - gapless with increasing field.   

Figure 2(c) shows the results of $C_\mathrm{m}(T)$ in the field region above $H_\mathrm{C2}$. 
At high temperatures, 
the broad maximum in $C_\mathrm{m}(T)$ shifts to higher temperatures as the field increases,  
while at low temperatures, 
the shoulder in $C_\mathrm{m}(T)$ seems to change into 
a cusp-like anomaly (hereafter referred to as a 2nd peak) due to the FIMO. 
For a closer look at the 2nd peak, 
we plot $C_\mathrm{m}(T)$ at around 1.5 K in this field region 
in the inset of Fig. 2(c), focusing on development of the 2nd peak due to the FIMO.
In the field range between 5 T and 6 T, 
a round peak shifts slightly to lower temperatures and develops gradually. 
At higher fields 7 T and 8 T, $C_\mathrm{m}(T)$ shows the clear anomaly.


\begin{figure}[bp]
\begin{center}
\includegraphics*[width=8cm]{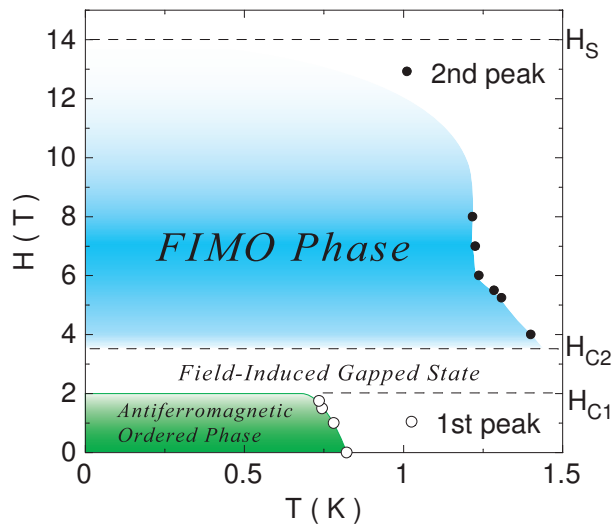}
\caption{Magnetic field versus temperature phase diagram of DMACuCl$_3$.}
\end{center}
\end{figure}

Plotting peak temperatures of the 1st and 2nd peaks in $C_\mathrm{m}(T)$, 
we show the $H$-$T$ phase diagram of DMACuCl$_3$ in Fig. 3. 
This phase diagram is considerably curious in two points as follows. 
One is the unusual field dependence of the spontaneous AF ordering temperature 
(shown as the open circles in Fig. 3) in the low field range below $H_\mathrm{C1}$. 
Generally, AF ordered phases have $H$-$T$ phase boundaries 
which would intersect the $H$-axis at the saturation field. \cite{af} 
Because spins are forced to align to the direction in parallel with the external field 
which suppresses the AF correlations. 
In contrast with this expectation, in the present case, 
$T_\mathrm{N}(H)$ changes rather mildly up to $H_\mathrm{C1}=2$ T and 
the $H$-$T$ phase boundary for the spontaneous AF ordering 
does not intersect the $H$-axis at $H_\mathrm{C1}$ 
where the antiferromagnetically ordered magnetic moments are partially saturated in the 1/2 magnetization plateau state. 
This fact indicates that this AF ordered phase involves other freedom of spin fluctuations. 
In our scenario, this phase is regarded as 
a partially ordered antiferromagnetic phase 
in view of that the only F dimers spontaneously ordered antiferromagnetically 
while AF dimers are essentially in the $S=0$ ground state still keeping strong quantum fluctuations. 
It should be noted that the AF dimers participate in the zero field ordering 
only by playing an important role to bring the AF correlation between the F dimers 
through the effective weak AF interaction mediated by intermediate singlet AF dimers 
as in the case of Diamond chain system Cu$_3$(CO$_3$)$_2$(OH)$_2$. \cite{az} 

The other distinguished anomaly to be noted is the unconventional phase boundary of the FIMO observed 
above $H_\mathrm{C2}$, as shown by the filled circles in Fig. 3. 
As is well known, a FIMO in spin-gapped systems is caused by 
a collapse of the energy gap due to Zeeman energy gain, 
by which a higher-lying excitation level is lowered and mixed into the ground state. 
In the usual case, the order parameter of the FIMO is the transverse staggered moment. 
Therefore, owing to its characteristic field dependence, 
the resultant FIMO phase has the reentrant $H$-$T$ phase boundary with a parabolic shape. 
As for the present case, a field-induced gapped state is realized between $H_\mathrm{C1}$ and $H_\mathrm{C2}$ as mentioned before. 
Then, in DMACuCl$_3$, a reentrant FIMO phase would be also expected to appear in the field range between $H_\mathrm{C2}$ and $H_\mathrm{S}$. 
In fact, 
the FIMO in DMACuCl$_3$ suddenly appears at $H_\mathrm{C2}$ as expected but, unexpectedly, 
at the finite temperature $T_\mathrm{N}(H_\mathrm{C2})$ and shows the opposite field dependence of $T_\mathrm{N}(H)$, 
in contrast with the conventional case where the phase boundary starts to appear from the zero temperature and $T_\mathrm{N}(H)$ 
increases with increasing field. \cite{Tl,K,NDMAP,ore} 
Even more interesting, $T_\mathrm{N}(H_\mathrm{C2})$ is much higher than $T_\mathrm{N}(0)=0.8$ K. 
This fact suggests that the intrachain correlation is much developed at $H_\mathrm{C2}$ than at zero field 
since it is generally considered that the 3D ordering sets in at the temperature 
$T_\mathrm{N}=(zJ'/k_\mathrm{B})\xi(T_\mathrm{N})$ 
where $z$, $J'$ and, $\xi(T_\mathrm{N})$ denote the number of interacting chains, 
the interchain interaction and the intrachain correlation length at $T_\mathrm{N}$ scaled by the lattice constant, respectively.

All such peculiarities which have never been observed in conventional spin-gapped systems 
can not be ascribed to any extrinsic natures of the system 
such as broadening of the peak due to the use of the powder sample. 
Because, the 2nd peak is sharp enough even at our highest field 8 T. 
Instead, we believe that these peculiarities originate most likely from the interplay between coexisting F and AF dimers. 
Here, 
we emphasize that the spontaneous ordering at zero field is basically concerned 
with the F dimers and the FIMO above $H_\mathrm{C2}$ is with the AF dimers. 
In consequence of the coexistence, 
the effective intrachain interaction is supposed to be field-dependent.
It is naturally conceivable from our scenario that the effective intrachain interaction 
between the AF dimers at $H_\mathrm{C2}$ is much stronger than that between the F dimers at zero field, 
because the former is mediated by intermediate fully-polarized F dimers 
while the latter is by intermediate singlet AF dimers. 
Finally, we discuss about the origin of the double-structure in $C_\mathrm{m}(T)$ in the 1/2 plateau field range. 
Apart from the distinguished sharp peaks due to both the spontaneous ordering and FIMO which are attributable to the effect of 
interchain interactions, 
we confirmed the overall nonsingular behaviors of $C_\mathrm{m}(T)$ 
can be reproduced qualitatively by the simple sum of two Schottky-type anomalies due to F and AF dimers as in zero field. \cite{ajiro} 
In particular, the calculated result features very well 
the field-dependent behaivor of the double-structure in $C_\mathrm{m}(T)$. 
The contribution from excitations in the F dimer with the Zeeman gap is 
the origin of the shoulder in the 1/2 plateau field range and shows a rather sensitive behavior to the field. 
On the other hand, the contribution of the gapped AF excitations in the AF dimers is 
the origin of the broad maximum and is a relatively insensitive to the field. 
At the same time, it is also apparent that the interplay between coexisting F and AF dimers can not be disregarded 
to explain the accurate shape of $C_\mathrm{m}(T)$. 
However, it is now clear that the observed double-structure is caused by the dual character of the low-lying excitations 
due to the coexistence of F and AF dimers in the present weakly coupled F-AF dimers chain system.

The observed coexistence of F and AF excitations is basically 
related to quantum ferrimagnetic chain systems with mixed spins, for which the coexistence 
of the gapless F and gapped AF excitations are predicted. \cite{ferri0, ferri1, ferri2}
In fields, the coexistence of the two kinds of excitations in quantum ferrimagnetic chain systems leads 
a double-structure in $C_\mathrm{m}(T)$, 
in which a lower-temperature and a high-temperature peaks are due to F and AF excitations, respectively. \cite{ferri2} 
As the field increases, two peaks merge into one below the field of an upper edge of the magnetization plateau, 
and the double-structure appears again above the 1/2 plateau field range. 
This field dependence of the double-structure in $C_\mathrm{m}(T)$ of the ferrimagnetic chain models
is qualitatively the same as the present case. 
In particular, 
the realized magnetic state in the 1/2 plateau field range is considered 
as a ferrimagnetic chain consisting of two sublattice moments, 
parallel (polarized) $S=1$ dimers and $S=0$ dimers connected alternatively in the 1D chain, 
generating the net magnetization as generally seen in the conventional ferrimagnetic system with different spins.

While, the above two models, isolated two kinds of dimers and ferrimagnetic chain ones, 
do not reproduce the spontaneous ordering and FIMO as a matter of course. 
This strongly indicates that the intervening interaction $J$ is essential to explain physical properties of DMACuCl$_3$ 
in addition to the remarkable coexistence of F and AF excitations. 

In conclusion, 
we performed the specific heat measurement on an $S=1/2$ spin-gapped system DMACuCl$_3$ in a magnetic field. 
The characteristic double-structure in $C_\mathrm{m}(T)$ was observed indicating coexistence of F and AF excitations 
in the 1/2 plateau field range. 
Further, two kinds of ordered phases, spontaneous AF and FIMO phases were found below and the above the plateau field range. 
$H$-$T$ phase boundaries for both ordered phases are substantially different from those of the other spin-gapped systems 
investigated so far. 
These behaviors were discussed in connection with the remarkable coexistence of F and AF excitations in a novel quantum 
spin model, constituting weakly coupled F and AF dimers.  

We would like express our sincere thanks to 
M. Matsumoto, S. Miyashita, N. Maeshima, and A. Tanaka for helpful comments and valuable discussions.  


\end{document}